# A Novel Hybrid Islanding Detection Method for Inverter-based DG

Maziar Isapour Chehardeh, Student member, IEEE

isapour@siu.edu

Ehsan. M. Siavashi, Student member, IEEE

e3mohamm@uwaterloo.ca

*Abstract—* A novel method for achieving a better performance using the combination of the available passive and active methods has been proposed. The algorithm detects the islanding in proper time by using harmonic detection, average rate of change of voltage and shifting power generation. Harmonic detection in this method decreases process time and also differentiates between islanding and other power systems events. For harmonic detection, extended Kalman filter has been used. Besides, reliability of the method increases using average rate of change of the voltage. The proposed method uses a strategy for decreasing the non-detection zone. In this strategy, minimum and maximum average rates of change of voltage limits are defined to improve the security of the system. Therefore, three main specifications of a proper method, reliability, security and time of process are achievable by combination of these passive and active methods. By applying different power system events under different power conditions, the proposed method has been verified in Simulink software.

I. INTRODUCTION

Nowadays, Renewable energy has been a research area of exponentially increasing interest, due to the attention of human beings towards the environment and energy problems. So, the electric power utilities are concerned with distributed generators including wind farm, photovoltaic, fuel cells, and micro sized grids as many good substitutes to solve the environmental problems and to manage the arising energy crisis. Distributed generation (DG) may make a contribution to improve quality of power, minimize peak loads and eliminate the need for reserve margin [1]. But there are many issues to be resolved before distributed generation comes into picture in large scale around the globe. One of the important issues is DG islanding.

Islanding operation occurs if one or more distributed generation (DG) continues to energize a part of the grid after the connection to the rest of the system has been lost due to any fault upstream or it can be any other kind of disturbance. IEEE 929-1988 standard [2] requires the disconnection of DG once it is islanded and IEEE 1547-2003 standard [1] stipulates a maximum delay of 2 seconds for detection of an unintentional island and all DGs cea sing to energize the distribution system. Hence, it is essential to detect the islanding both quickly and accurately.

Till now, many techniques have been proposed for detecting islanding [3]-[13]. Islanding techniques can be classified as remote and local techniques. Remote islanding techniques are based on PLCC systems and SCADA based anti-islanding systems [5][6]. Local techniques are based on the information and data available at the DG site. Local techniques are further classified as passive and active islanding techniques. Both techniques are basically based on variation of some parameters like voltage, frequency, harmonics etc.

Passive islanding detection techniques are based on the parameter measurement and setting up of threshold values for the measurable parameters. Passive methods continuously monitor and measure the system parameters like voltage magnitude, phase displacement, frequency etc. [7]-[10]. The variation in magnitude of the system parameters is compared with the tripping thresholds. Setting up of a proper threshold can help to differentiate between islanding and a grid connected condition. However the main challenge in passive methods is the appropriate selection of the tripping thresholds [11]-[13]. The rate of change of output power [5], [6], rate of change of frequency [7], rate of change of frequency over power [8], change of source impedance [9], harmonic detection [10][14][15][16] are some of the examples of passive techniques of islanding. These methods depend on the condition of load. Another problem is the setting of values of thresholds. Therefore, the passive techniques suffer from three main drawbacks: 1) the choice of suitable threshold; 2) a large non-detection zone (NDZ); 3) islanding detection is hard when the load and generation in the islanded system are closely matched.

In the active islanding techniques perturbations are injected locally into the test system and the responses of these perturbations are used in detecting the islanding condition in the distribution networks [11]-[13]. Slip- mode frequency shift method (SMS) [9], Phase shift methods for inverter based DG's [12][13][17], Reactive power export error detection and adaptive logic phase shift [7] are some of the active islanding detection techniques. However these techniques have some drawbacks. These techniques introduce perturbations into the test system, the detection time is very slow as a result extra time needed to analyze the system response of the perturbations. Moreover, these perturbations may break the power balance between the local load and the DG. Therefore, these techniques may not be effective in multi-DG power systems. Many other problems are associated with these techniques like injecting a distorted waveform, introducing high frequency signals. All these problems result in lower quality of output power. Therefore, there is a need to evolve an efficient methodology for detecting islanding of the distribution systems in the presence of distributed generation.

In this paper three objectives, security, reliability and fast detecting are considered. Therefore, by combination of the harmonic detection, rate of change of voltage and shifting power of DG, a novel method for achieving a better performance has been proposed. By this combination both advantage of active and passive methods are achievable. Moreover, in this method, it is not required to inject harmonic for observing the state of the system. A novel method is proposed to decrease detection time and NDZ.

## II. PROPOSED METHOD

Fig. 1 shows the flowchart of the proposed method. In the proposed method, two passive methods and one active method has been used. The Detection of specific harmonic and rate of change of voltage are the passive methods and real power shift is the active method used to detect islanding. As the amplitude of harmonics is amplified after islanding, for the first step, amplitude of specific harmonic is estimated by Extended Kalman Filter (EKF). Then, for differentiating between Islanding and other events in power systems like faults and load change the average rate of change of voltage (ARCV) is calculated for a specific time period. If these two parameters satisfy the conditional statement, for the detection of Islanding, the power of one of the DG has been changed to re-measure ARCV to finalize the decision about the specific event to identify the type of event.

These three methods were utilized in the proposed methodology due to three major reasons: 1) During Islanding operation, all the harmonics are amplified; 2) During Islanding operation, the value of ARCV is more than value of ARCV in the normal condition; 3) During Islanding, ARCV and power variation is remarkable comparing to the other events. However, each of three methods is not useful individually due to increase in disturbance amplitude of harmonics. Also, in single phase or 3-phase fault the ARCV becomes more than normal condition. Therefore, using only the rate of change of voltage as the Islanding factor takes long time to identify the event. On the other hand, it is not possible to change the power of the DG in each cycle. Therefore, in this paper, for differentiating between single phase fault and normal condition, instead of using change of fundamental frequency, the amplitude of 75Hz is used. The change of fundamental frequency is a slow process and takes time. Here, ARCV for two cycles will be calculated. Since ARCV for faults is greater than Islanding, $ARCV_{max}$ is set 14 for removing the severe faults. Moreover, in final decision, the power of one of DGs is set lower than others to observe the new ARCV.

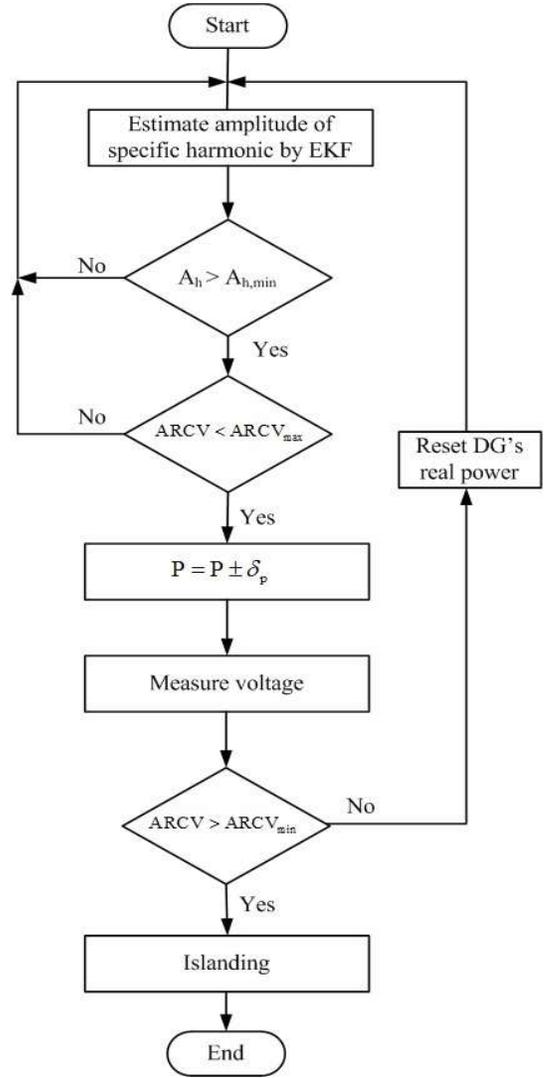

Fig. 1. Proposed method flowchart

For all events except Islanding, ARCV changes slightly so by setting $ARCV_{min} = 1$, Islanding will be detected.

Using harmonic detection results in less time for measuring the ARCV, therefore, the time of detection will be decreased using the harmonic detection. Besides, $ARCV_{min}$ and $ARCV_{max}$ decrease the non-detection zone. By this strategy, expectation is to reach the objectives such as reliability, security and inter-dependability.

## III. EXTENDED KALMAN FILTER

Consider a nonlinear system with state and measurement equations as

$$x_k = f_{k-1}(x_{k-1}, w_{k-1}) \quad (1)$$

$$z_k = h_k(x_k, v_k) \quad (2)$$

where $w_k$ and $v_k$ are independent white process and observation noises with probability distribution matrices $N(0,Q)$ and $N(0,R)$ respectively. Also $x_k$ and $z_k$ are the hidden state variables and the measurement respectively. Both $f(.)$ and $h(.)$ could be non-linear functions that are assumed as known functions.

Algorithm 1 illustrates The Extended Kalman filter where $z$ is the measurement vector, $\hat{x}_k^-$ and $P_k^-$ are the approximate state and covariance, $F$ and $H$ are the Jacobian matrices of partial derivatives of $f$ and $h$ with respect to $\hat{x}_k^-$. At each frame, the filter predicts the current state of the system and corrects this estimated state using measurement of the system. A Kalman gain ($K_k$) is computed to find the optimum feedback gain that minimizes the error covariance between the priory and the posteriori estimation. The updated estimate will then be used to predict the state for the next frame. Manner achievement of $F, H, W$ and $V$ is shown as below:

$$F = \frac{\delta f}{\delta x}(\hat{x}_{k-1}, u_k, 0) \quad (3)$$

$$W = \frac{\delta f}{\delta w}(\hat{x}_{k-1}, u_k, 0) \quad (4)$$

$$H = \frac{\delta h}{\delta x}(\tilde{x}_k, 0) \quad (5)$$

$$V = \frac{\delta h}{\delta v}(\tilde{x}_k, 0) \quad (6)$$

---

**Algorithm 1: Extended Kalman Filter**

**Step 1**: Predict the state with respect to initials:

$$\hat{x}_k^- = f(\hat{x}_{k-1}, 0)$$

**Step 2**: Compute the error covariance:

$$P_k^- = F_k P_{k-1} F_k^T + W_k Q_{k-1} W_k^T$$

**Step 3**: Compute the Kalman gain:

$$K_k = P_k^- H_k^T (H_k P_k^- H_k^T + V_k R_k V_k^T)$$

**Step 4**: Update the state estimate:

$$\hat{x}_k = \hat{x}_k^- + K_k (z_k - h(\hat{x}_k^-, 0))$$

**Step 5**: Update the error covariance

$$P = (I - K_k H_k) P_k^-$$

**Step 6**: Return to step 1

---

## IV. TEST SYSTEM

The test system which is used to show the effectiveness of the proposed method is shown in Fig. 2. The test network is a nine-bus system with the grid connected to the system through a circuit breaker. There are three inverter-based wind farms installed at bus numbers 8, 10 and 11. All wind DGs are inverter-based distributed generations instead of machine-based; each has specific control system to control the parameters for better efficiency of the system. The line data used in the test system as well as the load and generation data are provided in Table III.

## V. SIMULATION AND RESULTS

### A. Simulation

MATLAB Simulink is used in this paper to simulate the proposed method on the test system. Voltage base and power base are 12.7 KV and 10 MVA respectively.

In this paper, hysteresis controller method has been used to control the currents in the inverter block of DG. The main objective of the inverter control block is controlling the active power. For providing the reactive power, capacitor compensator is installed at bus 6.

For evaluating and verifying the proposed method, different events have been applied. In this paper, Islanding, decrement of load, 3-phase faults and Single-phase faults are considered for verification and evaluation. For the case of decreasing load, load at bus 4 (Load 2) is decreased to 10% of the normal condition. Also, 3-phase and single-phase faults are made to occur at bus 9. The single phase-fault is assumed to occur only for phase b.

The total simulation time is 0.3 sec and all events are applied at 0.1 sec. Also, for getting the proper response and measuring Average Rate of Change of Voltage (AVRC) real power is shifted i.e. power of DG 1 is decreased to 12% at 0.2 sec.

Observed voltage is assumed as below:

$$v(t) = \sum_{n=1}^{N} A_n \sin(n\omega t + \varphi_n) + A_{dc} e^{-\sigma t} + \varepsilon \quad (7)$$

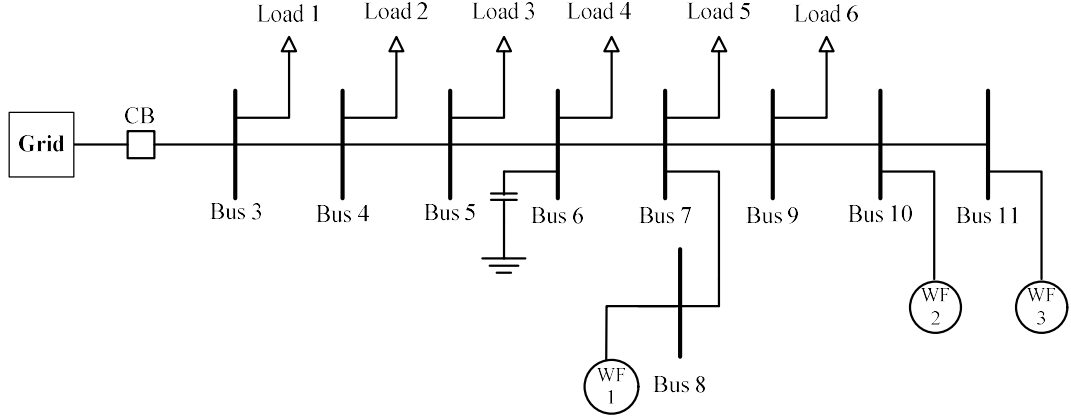

Fig. 2. Nine-bus test system

where
$A_n$    amplitude of the $n^{th}$ harmonic
$\phi_n$    phase of the $n^{th}$ harmonic
$n$    harmonic order ($N$ highest order)
$t = KT_s$    $T_s$ Sampling time and $K$ sampling instant
$\omega$    radian frequency
$A_{dc}$    amplitude of decay dc
$\varepsilon$    Noise

For detecting the harmonic frequency (75 Hz), variables for applying to Kalman filter are considered as below:

$$x_k(1) = e^{j\omega_1 T_s}$$
$$x_k(2) = a_1 e^{j(k\omega_1 T_s + \varphi_1)} \quad x_k(3) = a_1 e^{-j(k\omega_1 T_s + \varphi_1)}$$
$$x_k(4) = a_2 e^{j(2k\omega_1 T_s + \varphi_2)} \quad x_k(5) = a_2 e^{-j(2k\omega_1 T_s + \varphi_2)}$$
$$x_k(6) = a_3 e^{j(3k\omega_1 T_s + \varphi_3)} \quad x_k(7) = a_3 e^{-j(3k\omega_1 T_s + \varphi_3)}$$
$$x_k(8) = a_4 e^{j(4k\omega_1 T_s + \varphi_4)} \quad x_k(9) = a_4 e^{-j(4k\omega_4 T_s + \varphi_4)}$$
$$x_k(10) = a_5 e^{j(5k\omega_1 T_s + \varphi_5)} \quad x_k(11) = a_5 e^{-j(5k\omega_1 T_s + \varphi_5)}$$
$$x_k(12) = a_6 e^{j(6k\omega_1 T_s + \varphi_6)} \quad x_k(13) = a_6 e^{-j(6k\omega_1 T_s + \varphi_6)} \quad (8)$$
$$x_k(14) = a_7 e^{j(7k\omega_1 T_s + \varphi_7)} \quad x_k(15) = a_7 e^{-j(7k\omega_1 T_s + \varphi_7)}$$
$$x_k(16) = a_8 e^{j(8k\omega_1 T_s + \varphi_8)} \quad x_k(17) = a_8 e^{-j(8k\omega_1 T_s + \varphi_8)}$$
$$x_k(18) = a_9 e^{j(9k\omega_1 T_s + \varphi_9)} \quad x_k(19) = a_9 e^{-j(9k\omega_1 T_s + \varphi_9)}$$
$$x_k(20) = a_{5/4} e^{j((5/4)k\omega_1 T_s + \varphi_{5/4})}$$
$$x_k(21) = a_{5/4} e^{-j((5/4)k\omega_1 T_s + \varphi_{5/4})} \quad x_k(22) = K e^{-\sigma T_s}$$

By considering state variables, $f(x_k)$ and $h(x_k)$ in (1) and (2) are written as follows

$$f(x_k) = \begin{bmatrix} x_k(1) \;\; x_k(1)x_k(2) \;\; \dfrac{x_k(3)}{x_k(1)} \;\; .... \\ ... \; x_k(4)(x_k(1))^2 \;\; \dfrac{x_k(5)}{(x_k(1))^2} \;\; .... \\ ... \;\; \dfrac{x_k(21)}{(x_k(1))^{10}} \;\; x_k(22)x_k(22) \end{bmatrix}_{1\times 10} \quad (9)$$

$$h = \begin{bmatrix} 0 \; -0.5i \; 0.5i \; -0.5i \; 0.5i \; -0.5i \; 0.5i \; -0.5i \; .... \\ ... \; 0.5i \; -0.5i \; 0.5i \; -0.5i \; 0.5i \; -0.5i \; 0.5i \; 1 \end{bmatrix}_{1\times 22} \quad (10)$$

### B. Results

Case 1: In this case, the Load 5 is 5 MW and 1.6 MVAR, and results in power deficiency of 5 MW and 12.2 MVAR between DGs and Loads in the Islanding system. The voltage of bus 8 has been shown in Fig. 3 for different power events. By considering Table I, $A_{75}$ for 3-phase fault and Islanding is higher than the min-limit, therefore, load decrement and single-phase fault are filtered by harmonic detection block. Both events, Islanding and 3-phase fault do not pass over ARCV$_{max}$ condition, but by changing the power of DG1, the ARCV of 3-fault fault will not satisfy ARCV$_{min}$ condition and as a result of conditional block, it is filtered. Therefore, Islanding is detected. The value of the harmonic amplitude, ARCV after events and ARCV after changing power are indicated in Table I. The time process of the harmonic detection is slightly greater than one cycle and ARCV is measured in two cycles. Therefore, Islanding is detected in less than four cycles.

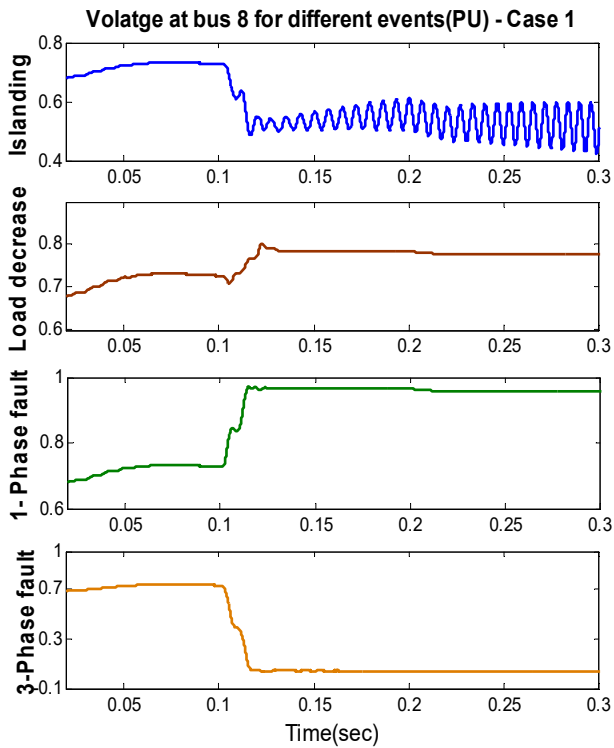

Fig. 3. Voltage of DG 1 at bus 8 under different events for case 1

Case 2: In this case, reactive power of Load 5 at bus 7 changes from 12.2 Mvar to 9.2 Mvar and active power is kept fixed. Also, active power of Load 2 at bus 4 changes from 7.3 MW to 5.3 MW. Therefore, in this case, power deficiency decreases to 2 MW and 2 Mvar. By applying different events, results are provided in Table I and the voltage of bus 8 has been shown in Fig. 4. Similar to the previous case, inter-harmonic (75 Hz) just for Islanding and 3-phase fault is amplified. After harmonics detection, ARCV block filters the 3-phase fault, because ARCV of 3-phase fault for this case is greater than $ARCV_{max}$. Fig. 4 reveals the change of voltage for different events. In islanding condition, the voltage at 0.1 sec and 0.2 sec is changed but for all other events, there is a slightly change in voltage at 0.2 sec. Therefore, Islanding is detected properly and efficiently.

Case3: In this case, change in active and reactive power of Load 5 at bus 7 from 5 MW to 3MW and 9.2 Mvar to 4 Mvar is applied, respectively, and the power deficiency becomes 1 MW and 1 Mvar. ARCV block only filters 3-phase fault because of its high ARCV value. In event of Islanding, voltages at bus number 8 responses to this change in power efficiently. Fig. 5 shows the result related to this case study.

Case 4: In this case, Load 5 is changed from previous case to 2 MW and 6.8 Mvar. Therefore, power deficiency changes to 0.1 MW and 0.1 Mvar. By considering the results of Table I and Fig. 6, it is obtained that similar to the previous cases, only 3-phase fault and Islanding pass through the harmonics detection block. Since 3-phase fault has ARCV value greater than threshold value, it filtered and just islanding is detected efficiently. Fig. 6 confirms the change of voltage of DG1 for different events.

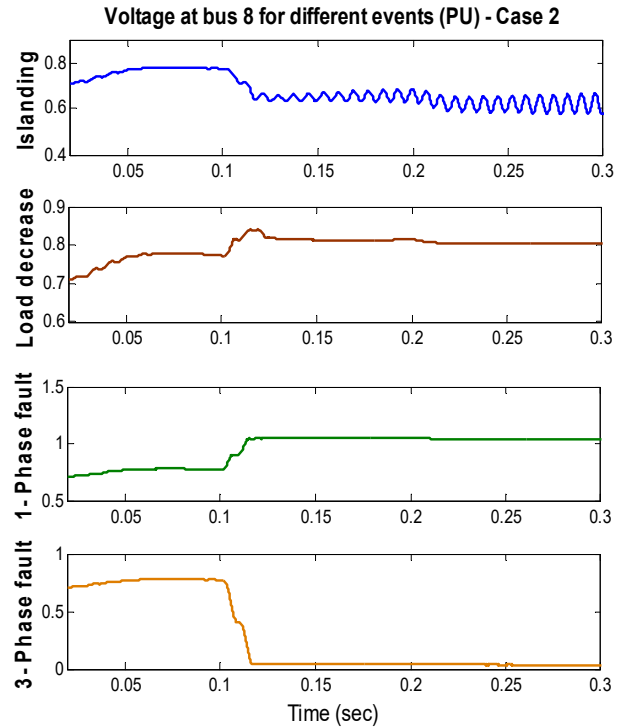

Fig. 4. Voltage of DG 1 at bus 8 under different events for case 2

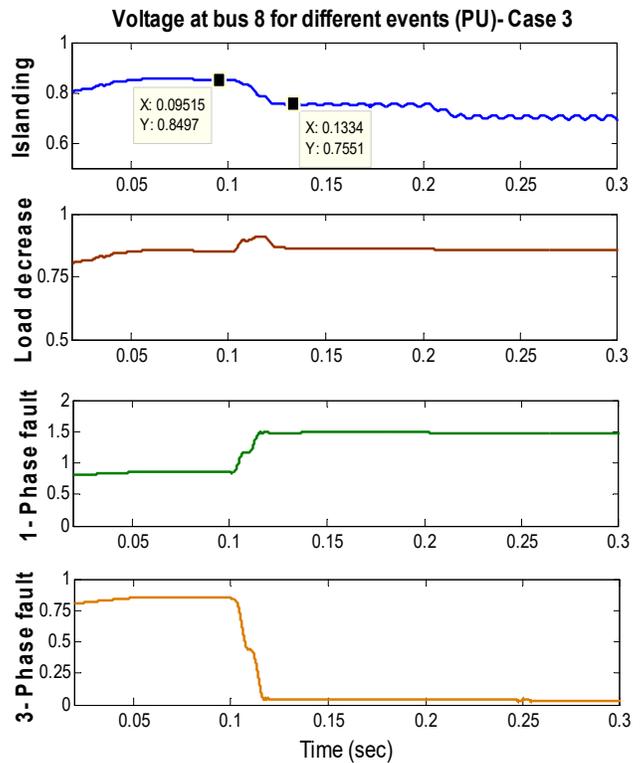

Fig. 5. Voltage of DG 1 at bus 8 under different events for case 2

TABLE I
Amplitude of harmonic and ARCV for different events

| Case 1 | | | |
|---|---|---|---|
| Event | A (75 Hz) pu | ARCV(1) | ARCV (2) |
| Islanding | 172.64 | 3.2215 | 2.9716 |
| 3-phase fault | 4.96 | 13.6943 | 0.0983 |
| 1-phase fault | 0.03 | 4.7750 | 0.1180 |
| Load decrease | 0.03 | 1.1457 | 0.0976 |
| Case 2 | | | |
| Event | A (75 Hz) pu | ARCV(1) | ARCV (2) |
| Islanding | 45.43 | 2.7285 | 1.4558 |
| 3-phase fault | 4.92 | 14.5979 | 0.0870 |
| 1-phase fault | 0.10 | 5.5911 | 0.2762 |
| Load decrease | 0.15 | 0.8069 | 0.1693 |
| Case 3 | | | |
| Event | A (75 Hz) pu | ARCV(1) | ARCV (2) |
| Islanding | 5.26 | 1.8058 | 1.0774 |
| 3-phase fault | 6.06 | 16.0325 | 0.1141 |
| 1-phase fault | 0.07 | 12.8253 | 0.1277 |
| Load decrease | 0.06 | 0.3187 | 0.0887 |
| Case 4 | | | |
| Event | A (75 Hz) pu | ARCV(1) | ARCV (2) |
| Islanding | 23.6 | 0.5561 | 1.0569 |
| 3-phase fault | 5.05 | 15.7184 | 0.0904 |
| 1-phase fault | 0.11 | 7.6998 | 0.2686 |
| Load decrease | 0.13 | 0.7422 | 0.1566 |

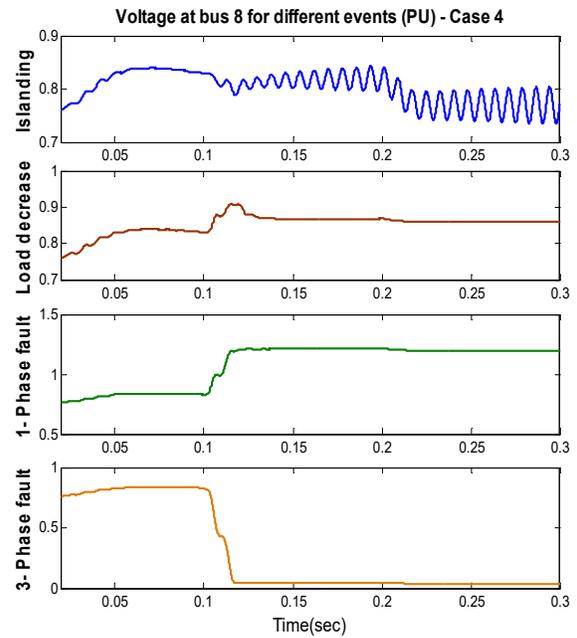

Fig. 6. Voltage of DG 1 at bus 8 under different events for case 4

TABLE II
Line data for the test system

| From Bus | To Bus | Resistance (Ω) | Reactance (mH) |
|---|---|---|---|
| 3 | 4 | 1.3825 | 2.62 |
| 4 | 5 | 0.18825 | 0.262 |
| 5 | 6 | 0.11295 | 0.1572 |
| 6 | 7 | 0.26355 | 0.3668 |
| 7 | 8 | 0.09036 | 0.12576 |
| 7 | 9 | 1.09185 | 1.5196 |
| 9 | 10 | 0.33885 | 0.4716 |
| 10 | 11 | 0.3765 | 0.524 |

TABLE III
Load and generation data for the test system

| Bus | PG (MW) | QG (MVAR) | PL(MW) | QL(Mvar) |
|---|---|---|---|---|
| 3 | 0 | 0 | 1.5 | 0 |
| 4 | 0 | 0 | 5.3 | 0 |
| 5 | 0 | 0 | 1 | 0 |
| 6 | 0 | 5 | 0.7 | 0 |
| 7 | 0 | 0 | 5 | 4 |
| 8 | 6 | 0 | 0 | 0 |
| 9 | 0 | 0 | 2 | 0 |
| 10 | 1.5 | 0 | 0 | 0 |
| 11 | 5 | 0 | 0 | 0 |

Paying more attention to the case 4, it is obvious that ARCV value for Islanding before changing the power is less because the value of mismatch of DGs and loads is very less. In this situation Islanding detection methods have problems and have poor effects on their performance. But in this method by considering harmonics detection and $ARCV_{max}$ as a conditioning factor, Non Detection Zone is decreased. Also, by adding harmonic detection block, ARCV requires less time to detect the islanding operation. Lastly, ARCV values obtained by the changing power, ensure Islanding detection and improve the overall reliability of the distribution network.

VI. CONCLUSION

In this paper, a novel method based on both passive and active Islanding detection methods has been proposed. For Islanding detection, at first step, amplitude of inter-harmonic is estimated by extended Kalman filter. Then average rate of change of voltage is measured. For decreasing the non-detection zone and enhancing the security, shifting power as an active method is used. Using harmonic detection, it is not necessary to consider more periods for calculating the rate of change of voltage. Therefore, the time of process will be decreased. Simulation results reveal that the proposed method met three main objectives, security, reliability and detection time.